\documentclass[aps,showpacs,twocolumn]{revtex4}
\usepackage{epsfig}
\usepackage{graphicx}
\usepackage{amsmath}
\usepackage{booktabs}

\begin{document}
\title{Can the state $Y(4626)$ be a $P$-wave tetraquark state $[cs][\bar{c}\bar{s}]$?}

\author{Chengrong Deng$^{a}{\footnote{crdeng@swu.edu.cn}}$,
        Hong Chen$^a{\footnote{chenh@swu.edu.cn}}$,
       and Jialun Ping$^b{\footnote{jlping@njnu.edu.cn}}$}

\affiliation{$^a$Department of Physics, Southwest University, Chongqing 400715, China}
\affiliation{$^b$Department of Physics, Nanjing Normal University, Nanjing 210097, China}

\begin{abstract} Stimulated by the state $Y(4626)$ recently reported by Belle Collaboration, we utilize a multiquark color flux-tube model with a multibody confinement potential
and one-glue-exchange interaction to make an exhaustive investigation on the diquark-antidiquark state $[cs][\bar{c}\bar{s}]$. Numerical results indicate that the appearance of the states $[cs][\bar{c}\bar{s}]$ like a dumb-bell, the larger the orbital excitation $L$, the more distinguished the shape. The mixing of the color configurations $\left[[cs]_{\bar{\mathbf{3}}_c}[\bar{c}\bar{s}]_{\mathbf{3}_c}\right]_{\mathbf{1}}$ and $\left[[cs]_{\mathbf{6}_c}[\bar{c}\bar{s}]_{\bar{\mathbf{6}}_c}\right]_{\mathbf{1}}$ in the ground states is strong while the color configuration $\left[[cs]_{\bar{\mathbf{3}}_c}[\bar{c}\bar{s}]_{\mathbf{3}_c}\right]_{\mathbf{1}}$ is absolutely predominant in the excited states. The main component of the state $Y(4626)$ can be interpreted as a $P$-wave state $[cs][\bar{c}\bar{s}]$. Its hidden-bottom partner is predicted in the model calculation. The states $X(4140)$, $X(4274)$, $X(4350)$, $X(4500)$ and $X(4700)$ are also discussed.

\end{abstract}


\maketitle

\section{Introduction}

The past decade or so has witnessed the great prosperity of the development of hadron physics. A large number of hidden charmed and bottomed hadrons were subsequently observed in experiments~\cite{review}, some of which, such as charged states $Z_b$ and $Z_c$, are difficult to be accommodated in the naive quark model. Very recently, Belle Collaboration reported a vector charmoniumlike state in the process of $e^+e^-\rightarrow D_s^+D_{s1}(2536)^-+c.c.$ via initial-state radiation~\cite{y4626}. The state has respectively a measured mass and width of $4265.9^{+6.2}_{-6.0}\pm0.4$ MeV and $49.8^{+13.9}_{-11.5}\pm4.0$ MeV and decays into a charmed antistrange and anticharmed-strange meson pair $D_s^+D_{s1}(2536)^-$ with a significance of 5.9 $\sigma$. The state is suggested as an exotic charmoniumlike state with $1^{--}$, called $Y(4626)$~\cite{y4626}, which provide an ideal opportunity to research the low-energy strong interaction. The most intuitive information provided by the decay behavior of the state $Y(4626)$ is that its main component is likely to be a tetraquark system $cs\bar{c}\bar{s}$.

In analogy with the deuteron, which is bound through the exchange of pion and other light mesons~\cite{meson-exchange}, Karliner and Rosner predicted the masses of tetraquark state $cs\bar{c}\bar{s}$ based on the proximity to thresholds of $D_s\bar{D}_s$ pairs~\cite{DsDs}. Inspired by the states $X(4140)$, $X(4274)$, $X(4500)$, $X(4700)$ and $Y(4140)$, the
tetraquark state $cs\bar{c}\bar{s}$ was also systematically researched in various theoretical framework, such as simple color-magnetic interaction models~\cite{cm1,cm2}, QCD sum rule~\cite{sumrule,sumrule2,sumrule3}, nonrelativistic and relativistic quark models~\cite{ncqm,rqm}, diquark model~\cite{diquark}, and lattice QCD~\cite{lqcdcscs}. A question then arises as to whether or not the main component of the state $Y(4626)$ can be described as the tetraquark state $cs\bar{c}\bar{s}$. Therefore, chiral constituent quark model was immediately used to describe the state $Y(4626)$ as a resonance state of $D_s^*D_{s1}(2536)$ with $1^-$~\cite{tp}.

A multiquark color flux-tube model based on the lattice QCD picture and the traditional quark models has been developed to study multiquark states, in which the multibody confinement potential is a dynamical mechanism in the formation and decay of the multiquark states~\cite{cftm}. Similar multibody string models were also extensively applied to study the properties of multiquark states~\cite{string1,string2}. In this work, we move on to the investigation on the tetraquark state $[cs][\bar{c}\bar{s}]$ to interpret the inner structure of the state $Y(4626)$ within the framework of the multiquark color flux-tube model, which is anticipated to exhibit new insights into the binding mechanisms in multiquark states, and maybe improve understanding of QCD in the nonperturbative regime.

This paper is organized as follows. After the introduction section, the presentation of the multiquark color flux-tube model is given in Sec. II. The wavefunction of the tetraquark state $[cs][\bar{c}\bar{s}]$ is shown in Sec. III. The numerical results and discussions are presented in Sec. IV. A brief summary is listed in the last section.

\section{Multiquark color flux-tube model}

Constituent quark models (CQM) are formulated under the assumption that the hadrons are color singlet nonrelativistic bound states of constituent quarks with phenomenological effective masses and interactions. One expects the dynamics of the CQM to be governed by QCD. The perturbative effect is well known one-gluon-exchange (OGE) interaction. The central part of the OGE interaction takes its form used extensively and is listed in the following~\cite{vijande},
\begin{eqnarray}
V_{ij}^{G} & = & {\frac{\alpha_{s}}{4}}\mathbf{\lambda}^c_{i}
\cdot\mathbf{\lambda}_{j}^c\left({\frac{1}{r_{ij}}}-
{\frac{2\pi\delta(\mathbf{r}_{ij})\mathbf{\sigma}_{i}\cdot
\mathbf{\sigma}_{j}}{3m_im_j}}\right), \nonumber
\end{eqnarray}
$\mathbf{\lambda}^c$ and $\mathbf{\sigma}$ respectively represent the Gell-Mann matrices and the Pauli matrices. The color-magnetic mechanism, which is proportional to the factor $\mathbf{\lambda}^c_{i}\cdot\mathbf{\lambda}^c_{j}\mathbf{\sigma}_{i}\cdot\mathbf{\sigma}_{j}$, in the OGE interaction leads to mass splitting among different color-spin configurations. $\alpha_s$ is a running strong coupling constant in the perturbative QCD~\cite{alphas},
\begin{equation}
\alpha_s(\mu^2)=\frac{1}{\beta_0\ln\frac{\mu^2}{\Lambda^2}},
\end{equation}
In this work, we takes the following form,
\begin{equation}
\alpha_s(\mu^2_{ij})=\frac{\alpha_0}{\ln\frac{\mu_{ij}^2}{\Lambda_0^2}},
\end{equation}
where $\mu_{ij}$ is the reduced mass of two interacting particles. The function $\delta(\mathbf{r}_{ij})$ should be regularized~\cite{weistein},
\begin{equation}
\delta(\mathbf{r}_{ij})=\frac{1}{4\pi r_{ij}r_0^2(\mu_{ij})}e^{-r_{ij}/r_0(\mu_{ij})},
\end{equation}
where $r_0(\mu_{ij})=\hat{r}_0/\mu_{ij}$. $\Lambda_0$, $\alpha_0$, $\mu_0$ and $\hat{r}_0$ are adjustable model parameters determined by fitting the data of $q\bar{q}$-mesons.

Color confinement is one of the most prominent features of QCD and should play an essential role in the low energy hadron physics. At present it is still impossible for us to
derive color confinement analytically from the QCD Lagrangian. Color confinement is a long distance behavior whose understanding continues to be a challenge in theoretical physics.
The color confinement potential in the traditional constituent quark model can be phenomenologically described as the sum of two-body interactions proportional to the
color charges and $r_{ij}^2$~\cite{isgur-karl},
\begin{eqnarray}
V^C&=& -a_c\sum_{i>j}^n\mathbf{\lambda}^c_{i}\cdot\mathbf{\lambda}^c_{j}r^2_{ij}
\end{eqnarray}
where $r_{ij}$ is the distance between two interacting quarks $q_i$ and $q_j$. The model can automatically prevent overall color singlet multiquark states disintegrating into several color subsystems by means of color confinement with an appropriate $SU_c(3)$ Casimir constant~\cite{drawback}. In contrast, the model allows a multiquark system
dissociating into color-singlet clusters, and it leads to interacting potentials within mesonlike $q\bar{q}$ and baryonlike $qqq$ subsystems in accord with the empirically known potentials~\cite{drawback}. However, the model is known to be flawed phenomenologically because it leads to power law van der Waals forces between color-singlet hadrons. In addition, it also leads to anticonfinement for symmetrical color structure in the multiquark system~\cite{anticonfinement}.

Up to now, color confinement can be established both from gauge-invariant lattice QCD (LQCD) simulations and from experimental observations like Regge trajectories~\cite{latticeguage,regge}. $q\bar{q}$ systems can be well reproduced at short distances by a linear potential. Such potential can be physically interpreted in a picture in which the quark and the antiquark are linked with a three-dimensional color flux tube. In the dual superconductor picture of color confinement~\cite{superconductor}, the color flux tube is formed due to the dual Meissner effect caused by monopole condensation. The chromoelectric field lines between color sources, like a quark and antiquark pair, are squeezed into a narrow flux tube along the line connecting the pair. Color flux tubes play significant roles in many interesting places of hadron physics, such as color confinement, quark pair creation and hadron structure.

LQCD calculations on baryons, tetraquark, and pentaquark states revealed that there exists flux-tube structures~\cite{lqcd}. In the case of a given spatial configuration of multiquark states, the confinement is a multibody interaction and can be simulated by a static potential which is proportional to the minimum of the total length of color flux-tubes. A naive flux-tube model, used in the present work, based on this picture has been constructed~\cite{cftm}. It takes into account multibody confinement with harmonic interaction approximation, i.e., where the length of the color flux-tube is replaced by the square of the length to simplify the numerical calculation. There are two theoretical arguments to support this approximation. One is that the spatial separations of the quarks (lengths of the color flux-tube) in hadrons are not large, so the difference between
the linear and quadratic forms is small and can be absorbed in the adjustable parameter, the stiffness. The other is that we are using a nonrelativistic description of the dynamics and, as was shown long ago~\cite{goldman}, an interaction energy that varies linearly with separation between fermions in a relativistic, first order differential dynamics has a wide region in which a harmonic approximation is valid for the second order (Feynman-Gell-Mann) reduction of the equations of motion. We calculated the $b\bar{b}$ spectrum by using quadratic and linear potentials, the results shown that the differences between two potentials are small for the low-lying states~\cite{SNChen}. In addition, the calculations on nucleon-nucleon interactions also support the replacement~\cite{difference}.

\begin{figure}
\resizebox{0.48\textwidth}{!}{\includegraphics{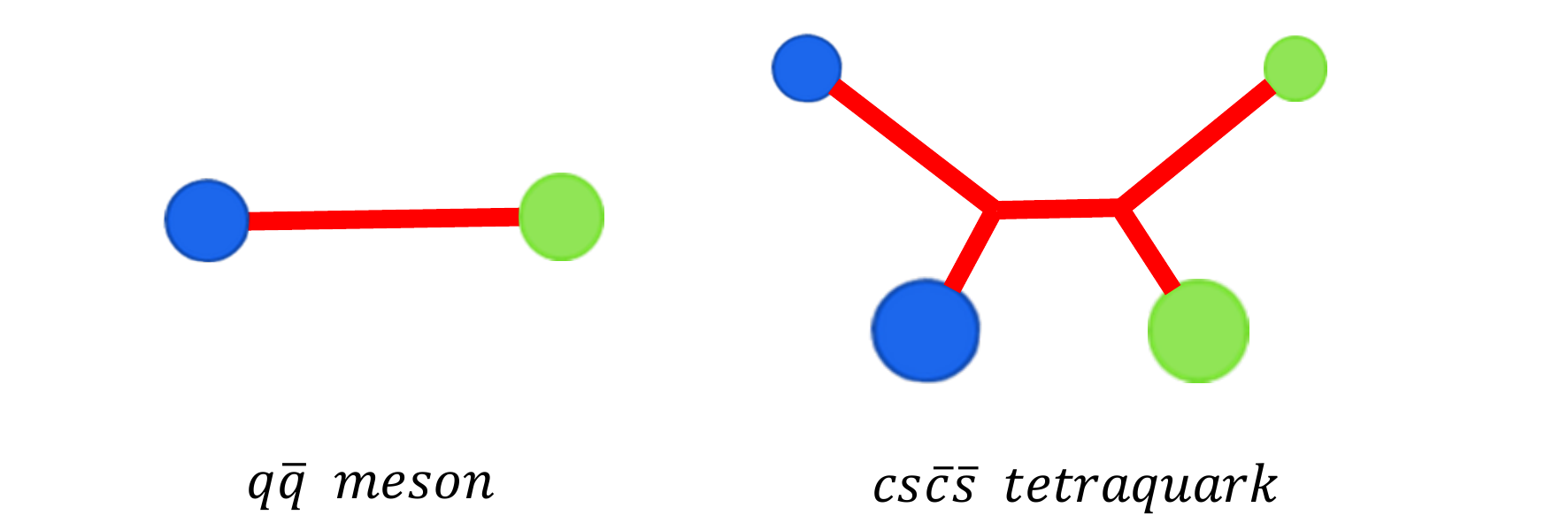}}
\caption{Color flux-tube structures.}
\label{spatial}
\end{figure}

The color flux-tube structures of $q\bar{q}$-mesons and the tetraquark state $[cs][\bar{c}\bar{s}]$ with diquark-antidiquark configuration are shown in Fig. 1, in which the blue and green disks respectively represent quark and antiquark. In the tetraquark state $[cs][\bar{c}\bar{s}]$, the big disk stands for a heavy quark while the small one stands for a light quark. The quark and antiquark in the mesons are linked with a three-dimensional color flux tube. A two-body confinement potential can be written as
\begin{eqnarray}
V_{min}^{C}(2)=Kr^2,
\end{eqnarray}
where $r$ is distance between the quark and antiquark and the parameter $K$ is the stiffnesses of a three-dimension color flux-tube and determined by fitting the ground heavy-meson spectra. In the state $[cs][\bar{c}\bar{s}]$, the codes of the quarks (antiquarks) $c$, $s$, $\bar{c}$ and $\bar{s}$ are assumed to be 1, 2, 3 and 4, respectively. According to a double-Y-shaped color flux-tube structure of the state $[cs][\bar{c}\bar{s}]$, a four-body quadratic confinement potential instead of linear one used in the LQCD can be written as,
\begin{eqnarray}
V^{C}(4)&=&K\left[ (\mathbf{r}_1-\mathbf{y}_{12})^2
+(\mathbf{r}_2-\mathbf{y}_{12})^2+(\mathbf{r}_{3}-\mathbf{y}_{34})^2\right. \nonumber \\
&+&
\left.(\mathbf{r}_4-\mathbf{y}_{34})^2+\kappa_d(\mathbf{y}_{12}-\mathbf{y}_{34})^2\right],
\end{eqnarray}
in which $\mathbf{r}_1$, $\mathbf{r}_2$, $\mathbf{r}_3$ and $\mathbf{r}_4$ respectively represent the position of the corresponding quark (antiquark). Two Y-shaped junctions $\mathbf{y}_{12}$ and $\mathbf{y}_{34}$ are variational parameters, which can be determined by taking the minimum of the confinement potential. The relative stiffness parameter $\kappa_{d}$ is $\kappa_{d}=\frac{C_{d}}{C_3}$~\cite{kappa}, where $C_{d}$ is the eigenvalue of the Casimir operator associated with the $SU(3)$ color representation $d$ at either end of the color flux-tube, such as $C_3=\frac{4}{3}$, $C_6=\frac{10}{3}$, and $C_8=3$.

The minimum of the confinement potential $V^C_{min}(4)$ can be obtained by taking the variation of $V^C(4)$ with respect to $\mathbf{y}_{12}$ and
$\mathbf{y}_{34}$, and it can be expressed as
\begin{eqnarray}
V^C_{min}(4)&=& K\left(\mathbf{R}_1^2+\mathbf{R}_2^2+
\frac{\kappa_{d}}{1+\kappa_{d}}\mathbf{R}_3^2\right),
\end{eqnarray}
The canonical coordinates $\mathbf{R}_i$ have the following forms,
\begin{eqnarray}
\mathbf{R}_{1} & = &
\frac{1}{\sqrt{2}}(\mathbf{r}_1-\mathbf{r}_2),~
\mathbf{R}_{2} =  \frac{1}{\sqrt{2}}(\mathbf{r}_3-\mathbf{r}_4), \nonumber \\
\mathbf{R}_{3} & = &\frac{1}{ \sqrt{4}}(\mathbf{r}_1+\mathbf{r}_2
-\mathbf{r}_3-\mathbf{r}_4), \\
\mathbf{R}_{4} & = &\frac{1}{ \sqrt{4}}(\mathbf{r}_1+\mathbf{r}_2
+\mathbf{r}_3+\mathbf{r}_4). \nonumber
\end{eqnarray}
The use of $V^C_{min}(n)$ can be understood here as that the gluon field readjusts immediately to its minimal configuration.

The diquark $[cs]$ and antidiquark $[\bar{c}\bar{s}]$ can be considered as compound bosons $\bar{Q}$ and $Q$ with no internal orbital excitation, and the orbital excitation $\mathbf{L}$ is assumed to occur only between $Q$ and $\bar{Q}$ in the present work. In order to facilitate numerical calculations, the spin-orbit interactions are assumed to take place approximately between compound bosons $\bar{Q}$ and $Q$, which is consistent with the work~\cite{spin-orbit}. The spin-orbit-related interactions can be expressed as follows
\begin{eqnarray}
V_{\bar{Q}Q}^{G,LS} & \approx& {\frac{\alpha_{s}}{4}}\mathbf{\lambda}^{\bar{c}}_{\bar{Q}}
\cdot\mathbf{\lambda}^c_{Q}{\frac{1}{8M_{\bar{Q}}M_{Q}}}\frac{3}{X^3}
\mathbf{L}\cdot\mathbf{S},\\
V_{\bar{Q}Q}^{C,LS} &\approx& \frac{K}{4M_{\bar{Q}}M_{Q}}\frac{\kappa_d}{1+\kappa_d}
\mathbf{L}\cdot\mathbf{S}.
\end{eqnarray}
where the masses of the compound bosons $M_{Q}=M_{\bar{Q}}\approx m_c+m_s$, $X$ is the distance between the two compound bosons, and $S$ stands for the total spin angular
momentum of the state $[cs][\bar{c}\bar{s}]$.

The completely Hamiltonian involving the multibody confinement potential and OGE interaction for the heavy mesons and the states $[cs][\bar{c}\bar{s}]$ can be presented as
\begin{eqnarray}
H_n & = & \sum_{i=1}^n \left(m_i+\frac{\mathbf{p}_i^2}{2m_i}\right)-T_{C}+\sum_{i>j}^n \left(V_{ij}^G+V_{ij}^{G,LS}\right) \nonumber\\
&+& V^C_{min}(n)+V^{C,LS}_{min}(n).
\end{eqnarray}
$T_{c}$ is the center-of-mass kinetic energy of the state and should be deducted; $\mathbf{p}_i$ is the momentum of the $i$-th quark (antiquark). LQCD computations on the
static tetraquark potential shown that the tetraquark potential is consistent with a four-body confining potential plus one gluon exchange Coulomb potentials~\cite{lqcdoge}.

It is worth mentioning that the multiquark color flux-tube model is not a completely new model but the updated version of the traditional CQM based on the color flux-tube picture
of hadrons in the LQCD. In fact, it merely modifies the two-body confinement potential in the traditional CQM into the multibody one to describe multiquark states with multibody interaction. Furthermore, the multiquark color flux-tube model can overcome the disadvantages of the traditional CQM.

\section{wavefunction of the state $[cs][\bar{c}\bar{s}]$}

The numerical results of the state $[cs][\bar{c}\bar{s}]$  should be solved using a complete wave function which includes all possible flavor-spin-color-spatial channels that contribute to a given well defined parity, isospin, and total angular momentum. Within the framework of the diquark-antidiquark configuration, the wave function of the state $[cs][\bar{c}\bar{s}]$ can be constructed as a sum of the following direct products of color $\chi_c$, isospin $\eta_i$, spin $\chi_s$ and spatial $\phi^G_{lm}$ terms
\begin{eqnarray}
\Phi^{[cs][\bar{c}\bar{s}]}_{IM_IJM_J} &=& \sum_{\alpha}\xi_{\alpha}\left[\left[\left[\phi_{l_am_a}^G(\mathbf{r})\chi_{s_a}\right]^{[cs]}_{s_a}
\left[\phi_{l_bm_b}^G(\mathbf{R})\right.\right.\right.\nonumber\\
&\times& \left.\left.\left.\chi_{s_b}\right]^{[\bar{c}\bar{s}]}_{s_b}\right ]_{S}^{[cs][\bar{c}\bar{s}]}
\phi^G_{LM}(\mathbf{X})\right]^{[cs][\bar{c}\bar{s}]}_{JM_J}~~~~~~~~~\label{wavefunction}\\
&\times& \left[\eta_{i_a}^{[cs]}\eta_{i_b}^{[\bar{c}\bar{s}]}\right]_{IM_I}^{[cs][\bar{c}\bar{s}]}
\left[\chi_{c_a}^{[cs]}\chi_{c_b}^{[\bar{c}\bar{s}]}\right]_{CW_C}^{[cs][\bar{c}\bar{s}]}
\nonumber
\end{eqnarray}
The subscripts $a$ and $b$ in the intermediate quantum numbers represent the diquark $[cs]$ and antidiquark $[\bar{c}\bar{s}]$, respectively. The summering index $\alpha$ stands for all possible flavor-spin-color-spatial intermediate quantum numbers. The parity of the state $[cs][\bar{c}\bar{s}]$ is related to the orbital excitations $\mathbf{L}$ as $P=(-1)^L$ because of $l_a=0$ and $l_b=0$. Considering a pair of charge-conjugated bosons $Q\bar{Q}$, we can obtain the $C$-parity $C=(-1)^{L+S-s_a-s_b}$ because the total wavefunction has to be completely symmetric under exchange of coordinates and spin of the bosons $Q$ and $\bar{Q}$.

The relative spatial coordinates $\mathbf{r}$, $\mathbf{R}$ and $\mathbf{X}$ in the state $[cs][\bar{c}\bar{s}]$ can be defined as,
\begin{eqnarray}
\mathbf{r}&=&\mathbf{r}_1-\mathbf{r}_2,~~~\mathbf{R}=\mathbf{r}_3-\mathbf{r}_4 \nonumber\\
\mathbf{X}&=&\frac{m_1\mathbf{r}_1+m_2\mathbf{r}_2}{m_1+m_2}-\frac{m_3\mathbf{r}_3+m_4\mathbf{r}_4}{m_3+m_4},\nonumber
%\mathbf{R}_c&=&\frac{m_1\mathbf{r}_1+m_2\mathbf{r}_2+m_3\mathbf{r}_3+m_4\mathbf{r}_4}{m_1+m_2+m_3+m_4}.\nonumber
\end{eqnarray}
In the dynamical calculation, the relative motion wave functions $\phi_{lm}^G$ can be expressed as the superposition of many different size Gaussian functions with well-defined quantum numbers as that of mesons in Sec. IV.

The color representation of the diquark $[cs]$ maybe antisymmetrical $\bar{\mathbf{3}}_c$ or symmetrical $\mathbf{6}_c$, whereas that of the antidiquark $[\bar{c}\bar{s}]$ maybe antisymmetrical $\mathbf{3}_c$ or symmetrical $\bar{\mathbf{6}}_c$. Coupling the diquark and the antidiquark into an overall color singlet according to color coupling rule only have two ways: $\left[[cs]_{\bar{\mathbf{3}}_c}\otimes[\bar{c}\bar{s}]_{\mathbf{3}_c}\right]_{\mathbf{1}}$ and $\left[[cs]_{\mathbf{6}_c}\otimes[\bar{c}\bar{s}]_{\bar{\mathbf{6}}_c}\right]_{\mathbf{1}}$. The spin of the diquark $[cs]$ is coupled to $s_a$ and that of the antidiquark  $[\bar{c}\bar{s}]$ to $s_b$. The total spin wave function of the state $[cs][\bar{c}\bar{s}]$ can be written as $S=s_a\oplus s_b$. Then we have
the following basis vectors as a function of the total spin $S$,
\begin{eqnarray}
S=\left\{
\begin{array}{ll}
\mbox{0,~~~$1\oplus1~\mbox{or}~0\oplus0$}   \\
\mbox{}\\
\mbox{1,~~~$1\oplus1,~1\oplus0~\mbox{or}~0\oplus1$}  \\
\mbox{}\\
\mbox{2,~~~$1\oplus1$}
\end{array}
\right. ,
\end{eqnarray}
For $S=0$ and 2, the state $[cs][\bar{c}\bar{s}]$ should have definite $C$-parity $(-1)^L$ because both the diquark and the antidiquark have the the same spin. For $S=1$,
the $C$-parity of the channel $1\oplus1$ is $(-1)^{L+1}$ while that of the channels $0\oplus1$ and $1\oplus0$ are $(-1)^L$.

The quarks $c$ and $s$ have isospin zero so that they do not contribute to the total isospin. The possible color-flavor-spin functions of the states $[cs][\bar{c}\bar{s}]$ with total spin $S$ can be written as,
\begin{eqnarray}
S=\left\{
\begin{array}{ll}
\mbox{0,~~~$\left[[cs]^{0,1}_{\bar{\mathbf{3}}_c}[\bar{c}\bar{s}]^{0,1}_{\mathbf{3}_c}\right]^0_{\mathbf{1}_c}$,
$\left[[cs]^{0,1}_{{\mathbf{6}}_c}[\bar{c}\bar{s}]^{0,1}_{\bar{\mathbf{6}}_c}\right]^0_{\mathbf{1}_c}$}   \\
\mbox{}\\
\mbox{1,~~~$\left[[cs]^{0,1}_{\bar{\mathbf{3}}_c}[\bar{c}\bar{s}]^{0,1}_{\mathbf{3}_c}\right]^1_{\mathbf{1}_c}$,
$\left[[cs]^0_{{\mathbf{6}}_c}[\bar{c}\bar{s}]^{0,1}_{\bar{\mathbf{6}}_c}\right]^{1}_{\mathbf{1}_c}$}  \\
\mbox{}\\
\mbox{2,~~~$\left[[cs]^1_{\bar{\mathbf{3}}_c}[\bar{c}\bar{s}]^1_{\mathbf{3}_c}\right]^2_{\mathbf{1}_c}$, $\left[[cs]^1_{\mathbf{6}_c}[\bar{c}\bar{s}]^1_{\bar{\mathbf{3}}_c}\right]^2_{\mathbf{1}_c}$}
\end{array}
\right. ,
\end{eqnarray}
where the superscript and subscript denote the spin and color representations, respectively. The number of the wave functions is big because the Pauli principle is out of operation in the state $[cs][\bar{c}\bar{s}]$.

\section{numerical calculations and analysis}

The starting point of the study on the state $[cs][\bar{c}\bar{s}]$ is to accommodate ordinary mesons in the multiquark color flux-tube model to determine model parameters. In order to avoid the misjudgement of the behavior of model dynamics due to inaccurate numerical results, a high precision numerical method is therefore indispensable. The Gaussian expansion method (GEM)~\cite{GEM}, which has been proven to be rather powerful to solve few-body problem in nuclear physics, is therefore widely used to study few-body systems. According to the GEM, the two-dody relative motion wave function can be written as,
\begin{eqnarray}
\phi^G_{lm}(\mathbf{r})=\sum_{n=1}^{n_{max}}c_{n}N_{nl}r^{l}e^{-\nu_{n}r^2}Y_{lm}(\hat{\mathbf{r}})
\end{eqnarray}
Gaussian size parameters are taken as geometric progression
\begin{eqnarray}
\nu_{n}=\frac{1}{r^2_n}, &r_n=r_1a^{n-1},
&a=\left(\frac{r_{n_{max}}}{r_1}\right)^{\frac{1}{n_{max}-1}}
\end{eqnarray}
The coefficient $c_n$ is determined by the dynamics of systems. With $r_1=0.3$~fm, $r_{n_{max}}=2.0$~fm and $n_{max}=7$, the converged numerical results can be arrived at.

The mass of $ud$-quark is taken to be one third of that of nucleon, other adjustable model parameters in Table \ref{para} can be determined by approximately strict solving twobody Schr\"{o}dinger equation to fit the masses of the ground states of heavy mesons in Table \ref{spectra}.
\begin{table}[ht]
\caption{Model parameters, quark mass and $\Lambda_0$ unit in MeV, $a_c$ unit in MeV$\cdot$fm$^{-2}$, $r_0$ unit in MeV$\cdot$fm and $\alpha_0$ is dimensionless.}\label{para}
\begin{tabular}{ccccccccccc}
\toprule[0.8pt] \noalign{\smallskip}
Para.   &~$m_{u,d}$~ & ~~$m_{s}$~~ & ~~$m_c$~~  &  ~~$m_b$~~   &   ~~$K$~~  &  ~~$\alpha_0$~~  & ~~$\Lambda_0$~~   & ~~$r_0$~~  \\
Valu.   &     313    &     494     &   1664     &    5006      &    $800$   &       4.25       &         40.85     &   119.3    \\
\toprule[0.8pt] \noalign{\smallskip}
\end{tabular}
\caption{Ground heavy-meson spectra, unit in MeV.} \label{spectra}
\begin{tabular}{ccccccccccccc}
\toprule[0.8pt] \noalign{\smallskip}
States         &~~~$D^{\pm}$~~~&~~$D^*$~~&~~~$D_s^{\pm}$~~~&~~$D_s^*$~~&~~~~$\eta_c$~~~~&~~$J/\Psi$~~&~~~$B^0$~~~      \\
Theo.          &    1886     &  2000   &   1982        &   2109    &   2965       &  3103      &   5261         \\
PDG.           &    1869     &  2007   &   1969        &   2112    &   2980       &  3097      &   5280         \\
\noalign{\smallskip}
\toprule[0.8pt]
\noalign{\smallskip}
States         &   $B^*$     & $B_s^0$ &   $B_s^*$     &   $B_c$   &   $B_c^*$    & $\eta_b$   &$\Upsilon(1S)$  \\
Theo.          &    5305     &  5346   &   5399        &   6244    &   6366       &  9376      &   9486         \\
PDG.           &    5325     &  5366   &   5416        &   6277    &    ...       &  9391      &   9460         \\
\toprule[0.8pt]
\end{tabular}
\end{table}

The mass spectrum of the state $[cs][\bar{c}\bar{s}]$ with $J^{PC}$ under the assumption of the total spin $S=0$, 1 and 2, and orbital excitation $L=0$, 1 and 2 in the multiquark color flux-tube model can be obtained by solving the four-body Schr\"{o}dinger equation with the well-defined trial wave functions of the state $[cs][\bar{c}\bar{s}]$ involving all possible channels,
\begin{eqnarray}
(H_4-E_4)\Phi^{[cs][\bar{c}\bar{s}]}_{IM_IJM_J}=0.
\end{eqnarray}
which are listed In Table \ref{cscs}. Using the wave function of the state $[cs][\bar{c}\bar{s}]$ obtained by solving the Schr\"{o}dinger equation, the mass and proportion of the color configurations $\left[[cs]_{\bar{\mathbf{3}}_c}[\bar{c}\bar{s}]_{\mathbf{3}_c}\right]_{\mathbf{1}}$ and  $\left[[cs]_{\mathbf{6}_c}[\bar{c}\bar{s}]_{\bar{\mathbf{6}}_c}\right]_{\mathbf{1}}$ can be arrived at and are given in Table \ref{cscs}. In the same way, the average distance
between any two particles and that between the diquark $[cs]$ and antiquark $[\bar{c}\bar{s}]$ can also be calculated and are shown in Table \ref{rms}.

The $\langle\mathbf{r}_{12}^2\rangle^{\frac{1}{2}}$ and $\langle\mathbf{r}_{34}^2\rangle^{\frac{1}{2}}$ respectively represent the size of the diquark $[cs]$ and antidiquark $[\bar{c}\bar{s}]$. It can be found from the Table \ref{rms} that they share the same value, around 0.6 fm, and are mainly determined by their own inner interactions. They are almost independent of the orbital excitation $L$ and are slightly influenced by the total spin $S$. The $\langle\mathbf{X}^2\rangle^{\frac{1}{2}}$ stands for the average distance between the diquark $[cs]$ and antidiquark $[\bar{c}\bar{s}]$, which greatly increases with the orbital excitation $L$. In the ground states, the short distance $\langle\mathbf{X}^2\rangle^{\frac{1}{2}}$ ranging from 0.35 fm to 0.41 fm is less than the size of the diquark $[cs]$ and antidiquark $[\bar{c}\bar{s}]$, which indicates that
the overlap of the two subclusters is extremely strong so that the picture of the diquark and antidiquark is not clear. In the excited states, the picture is gradually clear
because the diquark $[cs]$ and the dntiquark $[\bar{c}\bar{s}]$ are well separated with the increase of the orbital excitation $L$. The diagrammatic sketch of this picture is shown
in Fig. 2.

One can judge from the average distance in Table \ref{rms} that the diquark $[cs]$ and antidiquark $[\bar{c}\bar{s}]$ do not locate on a plane but twist into a three dimension spatial configuration, which is determined by the dynamics of the systems. Firstly, the configuration of the diquark $[cs]$ and antidiquark $[\bar{c}\bar{s}]$ is mainly determined by their inner dynamics. Under the condition of the specified $L$, the distance between the two subclusters is mainly determined by the competition between their relative motion and the confinement between the two subclusters because the former is inversely proportional to the distance while the latter is proportional to the distance. Secondly, the other interactions between the two subclusters result in the twist to arrive at a balance so that the diquark and antidiquark are not on a plane. The appearance of the tetraquark state $[cs][\bar{c}\bar{s}]$ like a dumb-bell, the larger the orbital excitation $L$, the more distinguished the shape, see Fig. 2. The multibody confinement potential, which
is a collective degree of freedom, based on the color flux-tube picture is the mainly dynamical mechanism of the formation of the picture. Lattice QCD calculation on the tetraquark states indicated that the three dimension spatial configuration is more stable than a planar one against transition into mesnons~\cite{twist}.

\begin{figure}
\resizebox{0.46\textwidth}{!}{\includegraphics{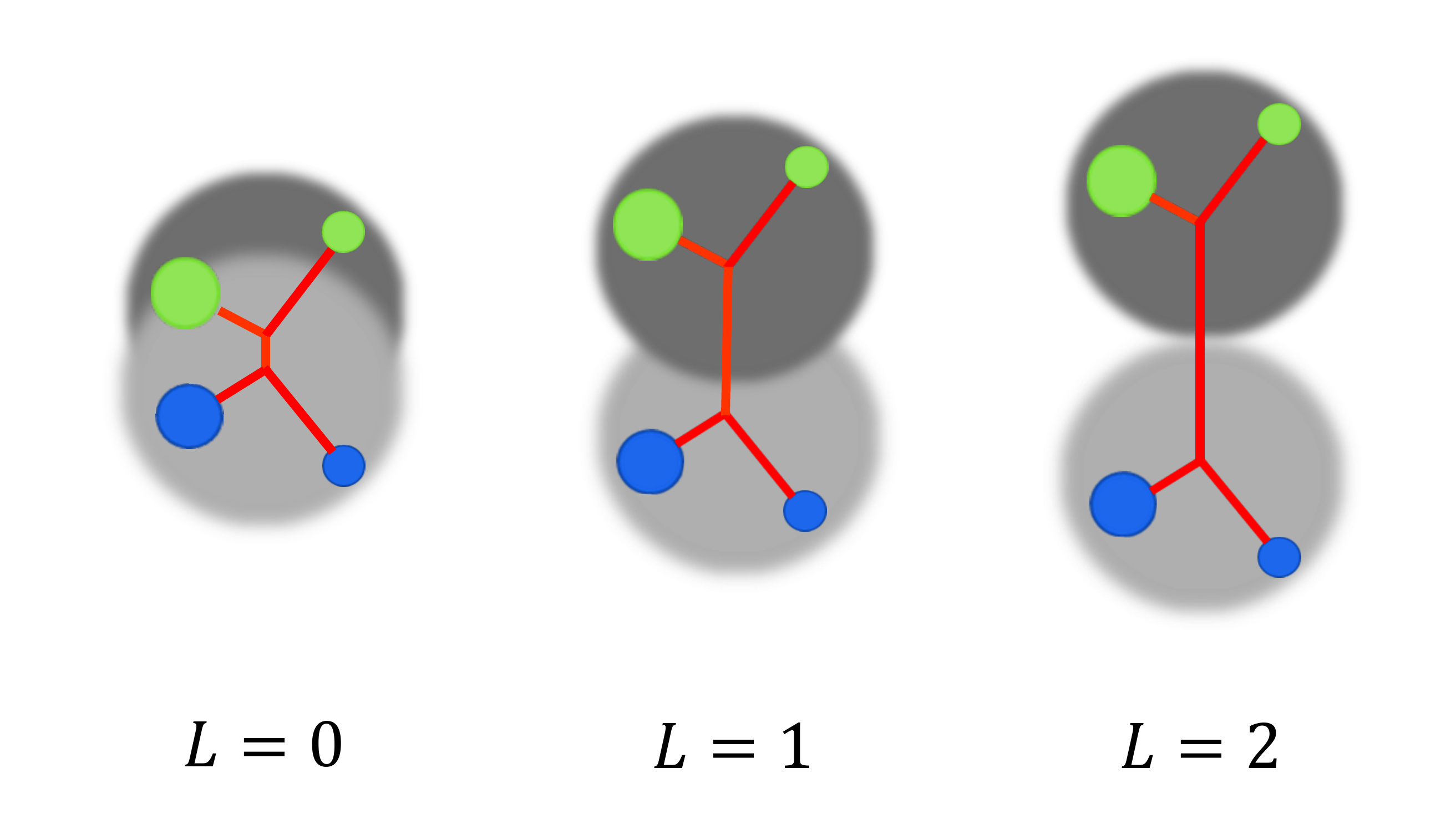}}
\caption{Diquark-antiquark picture.}
\label{spatial}
\end{figure}

Within the framework of diquark-antidiquark configuration, the state $[cs][\bar{c}\bar{s}]$ should be the mixture of the color configurations $\left[[cs]_{\bar{\mathbf{3}}_c}[\bar{c}\bar{s}]_{\mathbf{3}_c}\right]_{\mathbf{1}}$
and $\left[[cs]_{\mathbf{6}_c}[\bar{c}\bar{s}]_{\bar{\mathbf{6}}_c}\right]_{\mathbf{1}}$ by the coupling of color-related interactions. The mixing in the ground states is strong because the overlap of the diquark $[cs]$ and antidiquark $[\bar{c}\bar{s}]$ is extremely strong. Especially for the states with $0^{++}$ and $1^{+-}$, the color configuration $\left[[cs]_{\mathbf{6}_c}[\bar{c}\bar{s}]_{\bar{\mathbf{6}}_c}\right]_{\mathbf{1}}$ is predominant although the color configuration is not usually favored because of a repulsive interaction. In the excited states, the color configuration $\left[[cs]_{\bar{\mathbf{3}}_c}[\bar{c}\bar{s}]_{\mathbf{3}_c}\right]_{\mathbf{1}}$ is absolutely predominant so that the color configuration $\left[[cs]_{\mathbf{6}_c}[\bar{c}\bar{s}]_{\bar{\mathbf{6}}_c}\right]_{\mathbf{1}}$ can be completely ignored because the diquark $[cs]$ and antidiquark
$[\bar{c}\bar{s}]$ are well divided.

\begin{table*}
\caption{The mass spectra of the state $[cs][\bar{c}\bar{s}]$ with $J^{PC}$ in the color flux-tube model, C.C. represents the coupling results of the two color configurations,
unit in MeV.}\label{cscs}
\begin{tabular}{ccccccccccccccc}
\toprule[0.8pt]\noalign{\smallskip}
&$L=0$&&&&&$L=1$&&&&&$L=2$ \\
\toprule[0.8pt]\noalign{\smallskip}
$S$&~$J^{PC}$~&$\left[[cs]_{\bar{\mathbf{3}}_c}[\bar{c}\bar{s}]_{\mathbf{3}_c}\right]_{\mathbf{1}}$
&$\left[[cs]_{\mathbf{6}_c}[\bar{c}\bar{s}]_{\bar{\mathbf{6}}_c}\right]_{\mathbf{1}}$&C.C.&&~$J^{PC}$~& $\left[[cs]_{\bar{\mathbf{3}}_c}[\bar{c}\bar{s}]_{\mathbf{3}_c}\right]_{\mathbf{1}}$
&$\left[[cs]_{\mathbf{6}_c}[\bar{c}\bar{s}]_{\bar{\mathbf{6}}_c}\right]_{\mathbf{1}}$
&C.C.&&~$J^{PC}$~&  $\left[[cs]_{\bar{\mathbf{3}}_c}[\bar{c}\bar{s}]_{\mathbf{3}_c}\right]_{\mathbf{1}}$
&$\left[[cs]_{\mathbf{6}_c}[\bar{c}\bar{s}]_{\bar{\mathbf{6}}_c}\right]_{\mathbf{1}}$&C.C. \\
\noalign{\smallskip}
\toprule[0.8pt]\noalign{\smallskip}
$0$ &  $0^{++}$ & 4360,~$39.5\%$ & 4318,~$60.5\%$ & $4239$ && $1^{--}$ & 4629,~$95.2\%$ & 4802,~$4.8\%$ & 4620 && $2^{++}$ & 4868,~$99.1\%$ & 5163,~$0.9\%$ & 4865  \\                  \toprule[0.8pt]\noalign{\smallskip}
    &           &                &                &        && $0^{--}$ & 4665,~$97.4\%$ & 4875,~$2.6\%$ & 4659 && $1^{++}$ & 4899,~$99.1\%$ & 5188,~$0.9\%$ & 4897  \\
$1$ &  $1^{++}$ & 4378,~$64.1\%$ & 4416,~$35.9\%$ & $4330$ && $1^{--}$ & 4664,~$97.3\%$ & 4872,~$2.7\%$ & 4659 && $2^{++}$ & 4900,~$99.1\%$ & 5188,~$0.9\%$ & 4897  \\
    &           &                &                &        && $2^{--}$ & 4663,~$97.3\%$ & 4867,~$2.7\%$ & 4657 && $3^{++}$ & 4902,~$99.1\%$ & 5189,~$0.9\%$ & 4899  \\
\toprule[0.8pt]\noalign{\smallskip}
    &           &                &                &        && $0^{-+}$ & 4687,~$99.4\%$ & 4879,~$0.6\%$ & 4686 && $1^{+-}$ & 4925,~$99.7\%$ & 5164,~$0.3\%$ & 4924  \\
$1$ &  $1^{+-}$ & 4420,~$19.0\%$ & 4360,~$81.0\%$ & $4342$ && $1^{-+}$ & 4687,~$99.4\%$ & 4877,~$0.6\%$ & 4686 && $2^{+-}$ & 4926,~$99.7\%$ & 5165,~$0.3\%$ & 4925  \\
    &           &                &                &        && $2^{-+}$ & 4687,~$99.4\%$ & 4873,~$0.6\%$ & 4686 && $3^{+-}$ & 4928,~$99.7\%$ & 5166,~$0.3\%$ & 4927  \\
\toprule[0.8pt]\noalign{\smallskip}
    &           &                &                &        &&          &                &               &      && $0^{++}$ & 4935,~$99.6\%$ & 5190,~$0.4\%$ & 4934  \\
    &           &                &                &        && $1^{--}$ & 4705,~$99.2\%$ & 4911,~$0.8\%$ & 4704 && $1^{++}$ & 4935,~$99.6\%$ & 5190,~$0.4\%$ & 4934  \\
$2$ &  $2^{++}$ & 4430,~$75.8\%$ & 4455,~$24.2\%$ & $4418$ && $2^{--}$ & 4705,~$99.2\%$ & 4908,~$0.8\%$ & 4704 && $2^{++}$ & 4936,~$99.6\%$ & 5191,~$0.4\%$ & 4935  \\
    &           &                &                &        && $3^{--}$ & 4705,~$99.2\%$ & 4903,~$0.8\%$ & 4703 && $3^{++}$ & 4938,~$99.6\%$ & 5193,~$0.4\%$ & 4937  \\
    &           &                &                &        &&          &                &               &      && $4^{++}$ & 4940,~$99.6\%$ & 5194,~$0.4\%$ & 4939  \\
\toprule[0.8pt]\noalign{\smallskip}
\end{tabular}
%\end{table*}
%\begin{table*}
\caption{The average distance $\langle\mathbf{r}_{ij}^2\rangle^{\frac{1}{2}}$ between the $i$-th and $j$-th particle of the state $[cs][\bar{c}\bar{s}]$, unit in fm.}\label{rms}
\begin{tabular}{ccccccccccccccc}
\toprule[0.8pt] \noalign{\smallskip}
$~~S\oplus L$~~&~~$0\oplus0$~~&~~$0\oplus1$~~ &~~$0\oplus2$~~ &&~~$1\oplus0$~~&$1\oplus1$ &$1\oplus2$ &~~$1\oplus0$~~&$1\oplus1$ &$1\oplus2$ &~~$2\oplus0$~~&$2\oplus1$ &$2\oplus2$\\
\noalign{\smallskip}
$J^{PC}$& $0^{++}$& $1^{--}$& $2^{++}$ && $1^{++}$& $0,1,2^{--}$& $1,2,3^{++}$ &  $1^{+-}$& $0,1,2^{-+}$& $1,2,3^{+-}$ &  $2^{++}$& $1,2,3^{--}$& $0,1,2,3,4^{++}$ &   \\
\noalign{\smallskip}
\toprule[0.8pt] \noalign{\smallskip}
$\langle\mathbf{r}_{12}^2\rangle^{\frac{1}{2}}$  & 0.58 & 0.59 & 0.60 && 0.59 & 0.60 & 0.61 & 0.61 & 0.61 & 0.62 & 0.61 & 0.61 & 0.62   \\
$\langle\mathbf{r}_{34}^2\rangle^{\frac{1}{2}}$  & 0.58 & 0.59 & 0.60 && 0.59 & 0.60 & 0.61 & 0.61 & 0.61 & 0.62 & 0.61 & 0.61 & 0.62   \\
$\langle\mathbf{X}^2\rangle^{\frac{1}{2}}$       & 0.35 & 0.60 & 0.76 && 0.38 & 0.61 & 0.76 & 0.36 & 0.61 & 0.76 & 0.41 & 0.62 & 0.77   \\
$\langle\mathbf{r}_{13}^2\rangle^{\frac{1}{2}}$  & 0.40 & 0.63 & 0.79 && 0.43 & 0.64 & 0.79 & 0.40 & 0.64 & 0.79 & 0.45 & 0.65 & 0.80   \\
$\langle\mathbf{r}_{24}^2\rangle^{\frac{1}{2}}$  & 0.72 & 0.88 & 1.00 && 0.75 & 0.89 & 1.01 & 0.75 & 0.90 & 1.02 & 0.78 & 0.91 & 1.03   \\
$\langle\mathbf{r}_{14}^2\rangle^{\frac{1}{2}}$  & 0.58 & 0.77 & 0.90 && 0.61 & 0.78 & 0.91 & 0.60 & 0.78 & 0.91 & 0.64 & 0.79 & 0.92   \\
$\langle\mathbf{r}_{23}^2\rangle^{\frac{1}{2}}$  & 0.58 & 0.77 & 0.90 && 0.61 & 0.78 & 0.91 & 0.60 & 0.78 & 0.91 & 0.64 & 0.79 & 0.92   \\
\toprule[0.8pt]
\end{tabular}
\end{table*}

The color-magnetic interaction is a significant factor resulting in the mass splitting in the states with the same orbital excitation $L$ but different total spin $S$. For the ground states, the mass splitting between two adjacent states is about 90 MeV, see the states with $0^{++}$, $1^{++}$ and $2^{++}$ in Table \ref{cscs}. For the excited states with orbital excitations $L=1$ and $L=2$, the mass splitting are respectively around 40 MeV and 35 MeV, see the states in the lines 3, 4 and 6 in Table \ref{cscs}. The phenomenon of the stable mass difference between two adjacent states can be understood from the spatial distance shown in Table \ref{rms}, which is mainly determined by the orbital excitation $L$ but slightly influenced by the total spin $S$. The states with the total spin $S=1$ and opposite $C$-parity due to different spin-coupling models have close masses. The difference in the ground states is 12 MeV while the difference in excited states is less than 30 MeV.

The orbital excitation has a great influence on the mass of the state $[cs][\bar{c}\bar{s}]$. It induces a large mass splitting, about several hundred MeVs, among the states with different orbital angular momentum, which mainly comes from the kinetic energy and confinement potential because they both are proportional to the orbital excitation $L$. The spin-orbit interaction is extremely weak, which brings about a very small mass splitting, less than 5 MeV. Therefore, the masses of the excited states with the same $L$ and $S$ but different total angular momentum $J$ are almost degenerate, which is qualitatively consistent with the conclusion of the work~\cite{spin-orbit}. In addition, the tensor interactions is usually weak as the spin-orbit interaction~\cite{tension}, which is therefore frequently ignored in the preliminary research and should be taken into account in the further investigation of hyperfine structure.

The LHCb Collaboration recently confirmed the states $X(4140)$ and X(4274) in the $J/\Psi\phi$ invariant mass distribution and determined their spin-parity quantum numbers to be both $1^{++}$~\cite{4xstates1}, which has a large impact on its possible interpretations. The possibility of describing the state $X(4140)$ as $0^{++}$ or $2^{++}$ $D_s^{*+}D_s^{*-}$ molecule state was excluded~\cite{molecule}. At the same time, the depiction of the state $X(4274)$ as a molecular bound state or a cusp cannot accounts for its quantum numbers~\cite{4xstates1}. In the present work, the lowest energy of the state $[cs][\bar{c}\bar{s}]$ with $1^{++}$ is 4330 MeV, see Table \ref{cscs}, which is much higher, about 200 MeV, than the state $X(4140)$. In this way, it is difficult to accommodate the state $X(4140)$ as a state $[cs][\bar{c}\bar{s}]$ with $1^{++}$ in the multiquark color flux-tube model. However, the lowest energy is quite close to the mass of the state $X(4274)$, which implies a possibility that the main component of the state $X(4274)$ may be the state $[cs][\bar{c}\bar{s}]$ with $1^{++}$. Many of previous investigations on the two states also indicated that it is not easy to simultaneously arrange the two states within the same theoretical framework under the assumption of $1^{++}$~\cite{diquark,lqcdcscs,anisovich}. However, QCD sum rules and simple color-magnetic interaction models both can interpret the states $X(4140)$ and $X(4274)$ as $S$-wave states with $1^{++}$~\cite{cm1,cm2,sumrule2}.

Accompany with the states $X(4140)$ and $X(4274)$, the high $J/\Psi \phi$ mass region was investigated for the first time with good sensitivity and shows very significant structures, the states $X(4500)$ and $X(4700)$, which can be described as two $0^{++}$ resonances~\cite{4xstates1,4xstates2}. Comparing the dada, the mass of the lowest $S$-wave $[cs][\bar{c}\bar{s}]$ state with $0^{++}$ in the present work seems to be too light. It is therefore necessary to introduce radial excitation, $D$-wave or two $P$-wave angular excitation, which can satisfy the requirement of quantum numbers. The mass of the lowest $D$-wave $[cs][\bar{c}\bar{s}]$ state with $0^{++}$ in Table~\ref{cscs} is much higher, more than 200 MeV, than those of the states $X(4500)$ and $X(4700)$. Two $P$-wave excited states are higher than $D$-wave one. The two states should therefore not be the $D$-wave or two $P$-wave angular excited state $[cs][\bar{c}\bar{s}]$ in the present work. The masses of the fourth and fifth $S$-wave radial excited states $[cs][\bar{c}\bar{s}]$ are respectively 4466 MeV and $4699$ MeV, which can match with those of the states $X(4500)$ and $X(4700)$. Zhu also explained the two states as the radial excitation of $J^{P}=0^+$ tetraquark state~\cite{radial}. Chen et al interpreted the two states as the $D$-wave $[cs][\bar{c}\bar{s}]$ tetraquark states of $J^P=0^+$ within the framework of QCD sum rules~\cite{sumrule}. In addition to the $[cs][\bar{c}\bar{s}]$ explanation, the states $X(4500)$ and $X(4700)$ were described as conventional charmonium states with $4^3P_1$ and $5^3P_1$, respectively, in the nonrelativistic constituent quark model~\cite{p-wave}.

The Belle Collaboration observed a narrow $J/\Psi \phi$ peak at $4350.6^{+4.6}_{-5.1}\pm0.7$ MeV in two-photon collisions, which implies $J^{PC}=0^{++}$ or $2^{++}$~\cite{x4350}.
It is expected that the related experiments can provide more accurate information on the quantum numbers of the state in the future. If $J^{PC}=0^{++}$, one can find from Table \ref{cscs} that the pure $\left[[cs]_{\bar{\mathbf{3}}_c}[\bar{c}\bar{s}]_{\mathbf{3}_c}\right]_{\mathbf{1}}$ state with $0^{++}$ has a mass of $4360$ MeV, which is completely consistent with the experimental data. If $J^{PC}=2^{++}$, our prediction, 4418 MeV, is a little higher than the result reported by the experiment. Anyway, the state seems to be accommodated in the multiquark color flux-tube model just from the judgement of the mass and quantum number. In the simple color-magnetic interaction model, the states can be assigned as the state $[cs][\bar{c}\bar{s}]$ with $0^{++}$~\cite{cm2}. However, the state cannot be interpreted as a $[cs][\bar{c}\bar{s}]$ tetraquark with either $0^{++}$ and $2^{++}$ in the QCD sum rules~\cite{sumrule3}.

It can be found from Table~\ref{cscs} that the masses of the $P$-wave excited states $[cs][\bar{c}\bar{s}]$ with $1^{--}$ and total spin $S=0$, 1 and 2 are respectively 4620 MeV, 4659 MeV and 4704 MeV in the present work. The color configuration $\left[[cs]_{\bar{\mathbf{3}}_c}[\bar{c}\bar{s}]_{\mathbf{3}_c}\right]_{\mathbf{1}}$ in the three states is an overwhelming advantage so that the color configuration $\left[[cs]_{\mathbf{6}_c}[\bar{c}\bar{s}]_{\bar{\mathbf{6}}_c}\right]_{\mathbf{1}}$ can not be taken into account. The lowest state with the state $[cs][\bar{c}\bar{s}]$ with $1^{--}$ is most likely to be the best candidate of the main component of the state $Y(4626)$ because its mass completely consistent with that of the state $Y(4626)$. The masses of the other two $P$-wave states with $1^{--}$ are higher a little than that of the state $Y(4626)$, which also can be possile components of the state $Y(4626)$. As a mater of fact, the three states with $1^{--}$ can intermix through the tension interaction, which is left for precision calculation in the future. It can be anticipated that the tension interaction should be weak and does not change the present qualitative conclusion. As a byproduct, the masses of the hidden-bottom partners of the state $Y(4626)$ are estimated in the multiquark color flux-tube model, which are in the range of 11099 MeV to 11134 MeV. We propose to search for them in the $\Upsilon\phi$ invariant mass distribution in the future.

The assignment of the $[cs][\bar{c}\bar{s}]$ component of the states discussed in the present work is completed just based on the proximity to the experimental masses. The more rigorous test of the component of these states is to study their decay behavior. The states should eventually decay into several color singlet mesons due to their high energy. In the course of the decay, the three-dimension spatial structure must collapse first because of the breakdown of the color flux tubes, and then the decay products form by means of the recombination of color flux tubes. The decay widths are determined by the transition probability of the breakdown and recombination of color flux tubes, which is worthy of further research in the future work.

\section{summary}

We systematically study the state $[cs][\bar{c}\bar{s}]$ with diquark-antidiquark picture in the multiquark color flux-tube model with a multibody confinement potential and one-gluon-exchange interaction. The size of the diquark $[cs]$ and diquark $[\bar{c}\bar{s}]$ share the same value, which is almost independent of the orbital excitation $L$ and is slightly influenced by the total spin $S$. The average distance between the diquark $[cs]$ and diquark $[\bar{c}\bar{s}]$ greatly increases with the orbital excitation $L$. The appearance of the tetraquark state $[cs][\bar{c}\bar{s}]$ thereofre like a dumb-bell, the larger the orbital excitation $L$, the more distinguished the shape. The multibody confinement potential is the mainly dynamical mechanism of the formation of the picture. The mixing of the two color configurations $\left[[cs]_{\bar{\mathbf{3}}_c}[\bar{c}\bar{s}]_{\mathbf{3}_c}\right]_{\mathbf{1}}$ and $\left[[cs]_{\mathbf{6}_c}[\bar{c}\bar{s}]_{\bar{\mathbf{6}}_c}\right]_{\mathbf{1}}$ in the ground states is strong while the color configuration $\left[[cs]_{\bar{\mathbf{3}}_c}[\bar{c}\bar{s}]_{\mathbf{3}_c}\right]_{\mathbf{1}}$ is favored and absolutely predominant in the excited states.

The state $Y(4626)$ can be well interpreted as the $P$-wave state $[cs][\bar{c}\bar{s}]$ with $1^{--}$ in the multiquark color flux-tube model, its hidden bottom partner has a
mass in the range of 11099 MeV to 11134 MeV and can be searched for in the $\Upsilon\phi$ invariant mass distribution in the future. The properties of the states $X(4140)$, $X(4274)$, $X(4350)$, $X(4500)$ and $X(4700)$ are also discussed in the model. The state $X(4274)$ is quite close to the lowest mass of the state $[cs][\bar{c}\bar{s}]$ with $1^{++}$. The states $X(4500)$ and $X(4700)$ can be described as the fourth and fifth excited states of the ground state $[cs][\bar{c}\bar{s}]$ with $0^{++}$. The masses of the ground states $[cs][\bar{c}\bar{s}]$ with $0^{++}$ and $2^{++}$ are both close to that of the state $X(4350)$. However, the lightest state $X(4140)$ regarded as the state $[cs][\bar{c}\bar{s}]$ with $1^{++}$ can not be accommodated in the model. These results in some extent reinforce the validity of the multiquark color flux-tube model to quantitatively describe the phenomenology of the multiquark states and get insights on the dynamics that leads to their formation.

As an outlook of the continuation of this work, the string flip-flop potential regarded as the correct phenomenological model for the confinement should be taken into accounted. The flip-flop potential is important for the properties of the tetraquark states especially for the decay process into two mesons.

\acknowledgments

{This research is partly supported by the National Science Foundation of China under Contracts Nos. 11875226 and 11775118, the Natural Science Foundation of Chongqing, China under Project No. cstc2019jcyj-msxmX0409 and Fundamental Research Funds for the Central Universities under Contracts No. SWU118111.}

\end{document}